\documentclass[a4paper,11pt]{article}
\pdfoutput=1 
\usepackage{jinstpub} 
\usepackage{soul}
\usepackage{url} 
\usepackage{hyperref}
\usepackage{graphicx}

\usepackage{ulem} 
\usepackage{lineno}
\usepackage{multirow}
\usepackage{caption}
\usepackage{subcaption}
\usepackage[utf8]{inputenc}

\title{Additive manufacturing of inorganic scintillator-based particle detectors}
\author[a,1]{T.~Sibilieva,} 
\author[a]{V.~Alekseev,}
\author[b]{S.~Barsuk,}
\author[c,d,e]{S.~Berns,}
\author[c,d,e]{E.~Boillat,}
\author[a,b]{I.~Boiaryntseva,}
\author[a]{A.~Boyarintsev,}
\author[f,i]{A.~Carbone,}
\author[g]{A.~De Roeck,}
\author[g]{S.~Dolan,}
\author[a]{T.~Driuk,}
\author[h]{A.~Gendotti,}
\author[a]{I.~Gerasymov,}
\author[a]{B.~Grynyov,}
\author[c,d,e]{S.~Hugon,}
\author[h]{U.~Kose,}
\author[a]{O.~Opolonin,}
\author[h]{A.~Rubbia,}
\author[h]{D.~Sgalaberna,}
\author[a]{N.~Sibilyev,}
\author[a]{S.~Tretyak,}
\author[h]{T.~Weber,}
\author[h]{J.~Wuthrich,}
\author[h]{X.~Y.~Zhao}

\affiliation[a]{Institute for scintillation materials National Academy of Science of Ukraine (ISMA NAS of Ukraine), Nauki ave. 60, Kharkiv 61072, Ukraine}
\affiliation[b]{Laboratoire de Physique des 2 Infinis, Ir\'ene Joliot-Curie, Universit\`e Paris-Saclay,
Universit\`e de Paris, IN2P3/CNRS, 91405 Orsay, France}
\affiliation[c]{Haute Ecole Sp\'ecialis\'ee de Suisse Occidentale (HES-SO), CH-2800 Del\'emont, Route de Moutier 14, Switzerland}
\affiliation[d]{Haute Ecole d'Ing\'enierie du canton de Vaud (HEIG-VD), CH-1401 Yverdon-les-Bains, Route de Cheseaux 1, Switzerland}
\affiliation[e]{COMATEC-AddiPole, CH-1450 Sainte-Croix, Technopole de Sainte-Croix, Rue du Progr\`es 31, Switzerland}
\affiliation[f]{Istituto Nazionale Di Fisica Nucleare (INFN), Sezione di Bologna, Viale C. Berti Pichat, 6/2, 40127, Bologna, Italy}
\affiliation[g]{Experimental Physics department, European Organization for Nuclear Research (CERN), Esplanade des Particules 1, 1211 Geneva 23, Switzerland}
\affiliation[h]{Institute for Particle physics and Astrophysics, ETH Zurich, Otto-Stern-Weg 5, CH-8093 Zurich, Switzerland}
\affiliation[i]{Università di Bologna, Via Zamboni, 33, 40126 Bologna, Italy}

\emailAdd{sibilieva@isma.kharkov.ua}
\note{Corresponding author}

\abstract{Inorganic scintillators are widely used for scientific, industrial and medical applications. The development of 3D printing with inorganic scintillators would allow fast creation of detector prototypes for registration of ionizing radiation, such as alpha and beta, gamma particles in thin layers of active material and soft X-ray radiation. This article reports on the technical work and scientific achievements that aimed at developing a new inorganic scintillation filament to be used for the 3D printing of composite scintillator materials: study and definition of the scintillator composition; development of the methods for the inorganic scintillator filament production and further implementation in the available 3D printing technologies; study of impact of the different 3D printing modes on the material scintillation characteristics. Also, 3D printed scintillators can be used for creation of combined detectors for high-energy physics.}


\begin{document}
\maketitle
\flushbottom

\section{Introduction}
\label{sec:intro}

In the last years, more and more attention has been paid to additive manufacturing technologies (AM),
also called 3D printing, to obtain functional materials in the modern material science.
The improvement in the available AM technologies is accompanied by a significant reduction of the costs,
which makes 3D printing a viable solution not only for fast prototyping but also for the final
production of finite objects. 

The 3D printed Detector (3DET) R\&D collaboration~\cite{3DETwebsite} aims at developing 3D printing
for scintillator-based particle detectors. For instance, the 3DET collaboration aims to
3D print both the active scintillator part and the other detector components, such as dark box,
optical connectors, etc. As we have shown recently, plastic scintillators based on polystyrene
were manufactured using fused deposition modeling (FDM) 3D printing technology with characteristics
comparable with the standard production techniques such as cast polymerization, extrusion and injection
molding~\cite{Berns}. Moreover, FDM technology allows printing of both scintillator and reflector
at once by using two extruders~\cite{3Dmatrix}. Hence, it allows to create multi-element scintillation
detectors, such as an array of scintillation cubes or tiles separated by a reflector, as well as to create
combined scintillation detectors from several scintillation materials. 

Although the results obtained in~\cite{3Dmatrix} are very promising, work is in progress in order
to improve the 3D printed scintillator performance, the geometrical tolerance and the
uniformity in the reproducibility of multiple samples, towards the first real particle detector
ever 3D printed. The main goal will be the fast and cost-effective production of massive
fine-granularity polystyrene-based plastic-scintillator detectors with applications in
high-energy physics (HEP), such as neutrino active scintillator targets or particle calorimetry~\cite{Blondel}.
However, AM also shows a strong potential for applications beyond HEP. 

A large number of scintillation materials and detectors for high energy physics are known.
In addition to plastic scintillators, inorganic scintillators are widely used for radiation detection, such as
Gd$_{3}$Al$_{2}$Ga$_{3}$O$_{12}$:Ce (GAGG:Ce), Y$_{2}$SiO$_{5}$:Ce (YSO:Ce), Y$_{3}$Al$_{5}$O$_{12}$:Ce (YAG:Ce), Lu$_{3}$Al$_{5}$O$_{12}$:Ce (LuAG:Ce),
Gd$_{2}$O$_{2}$S:Pr (GOS:Pr), ZnSe:Al, CsI:Tl, etc. \cite{Dujardin,Auffray,Martinazzoli,Martinazzoli_2}.

The development of 3D printing with inorganic scintillators would allow to create fast prototypes
of detectors for registration of charged particles, such as alpha and beta, gamma 
particles in thin layers of active material and soft X-ray radiation. For 3D printing using
FDM technology, a thermoplastic filament is required as a starting material. The production
of such a filament for printing scintillators is possible if a composition is created from
granules of inorganic scintillators and transparent thermoplastic polymers. The implementation
of 3D printing in the manufacturing of composite inorganic scintillators makes the creation of
scintillators with the required geometry possible without additional machining. In fact,
all 3D printed samples presented in this work were not subjected to neither grinding or polishing. 

Previously~\cite{Gerasymov,Boyarintsev,Boyarntsev_2,Kerch,Nepokupnaya}, we investigated composite scintillators
based on grains of
inorganic single crystals. Polysiloxane elastomer was used as a binder.
Single crystals were mechanically grinded up to obtain scintillation grains.
After that, a set of calibrated sieves was used to select the fraction of grains with the optimal size.
Then, the grains were introduced in the polysiloxane elastomer and the composition was mixed.
Finally, the gel composition was introduced into a forming container, in which it was left to
complete its polymerization. As a result, the scintillator is obtained and can be taken from the
forming container. Composite scintillators with inorganic single crystals are a promising
alternative to bulk inorganic scintillation detectors. In particular, to make large area
detectors at a reasonable price thanks to the possibility of using the waste from the
production of single crystals.

This article reports on the technical work and scientific achievements that aimed at developing
a new inorganic scintillation filament to be used for the 3D printing of composite scintillator
materials: study and definition of the scintillator composition; development of the methods for
the inorganic scintillator filament production and further implementation in the available
3D printing technologies; study of impact of the different 3D printing modes on the material
scintillation characteristics. In the following sections we will discuss the results of latest R\&D.

\section{Obtaining scintillation filaments with inorganic granules and 3D printing of scintillators}
\label{sec:second}
In this work, inorganic scintillator based 3D printed samples from ZnSe:Al, GOS:Pr,  GAGG:Ce and CsI:Tl
were obtained and tested. Such inorganic scintillators are widely used in radiography and X-ray
computed tomography from medical to industrial applications due to their high efficiency X-ray
absorption and high light output. For instance, a thickness of 1~mm is sufficient to absorb
almost all the incident X-rays at a few tens of keV and produce a sensible signal.

The scintillator filaments were made on the basis of the above mentioned inorganic scintillators
with the addition of polymeric materials as optical binder. PS (polystyrene), ABS (acrylonitrile
butadiene styrene), SBS (styrene-butadiene-styrene) and PMMA (polymethyl methacrylate) polymer
pellets were chosen as optical binder due to their high transparency to visible light.

The use of inorganic scintillation granules as part of filament compositions significantly expands
the possibilities of using 3D printing technology also for inorganic scintillators. The material
for 3D printing with FDM technology is a thermoplastic filament, produced by mixing inorganic
granules with polymer pellets. 

\begin{figure}[h]
\centering
\includegraphics[width=11cm]{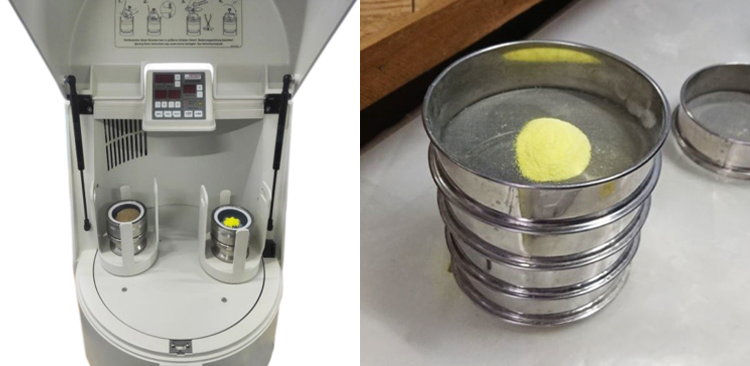}
    
\caption{\label{fig:figure1} Left: Planetary mill Fritsch Pulverisette 5/2; Right: set of sieves for fractioning of the inorganic scintillation granules}
\end{figure}

The first stage consists of the grinding of scintillation inorganic crystals.
As shown in Figure~\ref{fig:figure1}a,
the planetary mill Fritsch Puluerisette 5/2 was used for grinding and producing
inorganic scintillation granules. An appropriate set of sieves was used to select the
inorganic scintillation granules of the required size: 1 - 15, 45 - 63, 63 - 100 and 100 - 140~$\mu$m.

\begin{figure}[h]
\centering
{\includegraphics[width=14cm]{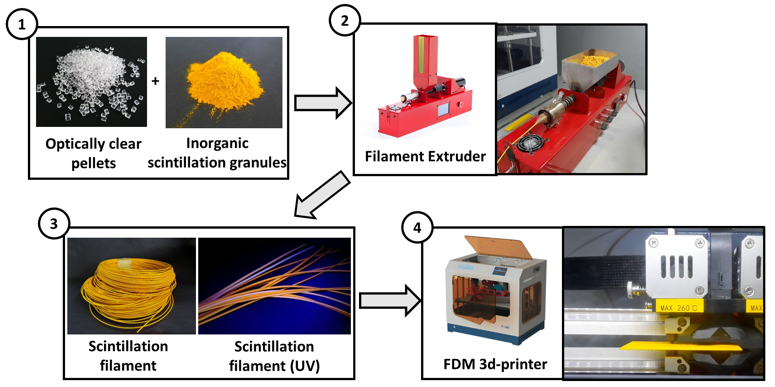}}
\caption{\label{fig:figure2} The process of production of filament with inorganic scintillation granules and 3D printing of the scintillator.}
\end{figure}

The next step is the production of scintillation filament for 3D printing.
As shown in Figure~\ref{fig:figure2}, its production consists of the following steps:
\begin{itemize}

\item plasticizer and scintillation inorganic granules were added to the optically-transparent polymer pellets (binder) and mixed;

\item the resulting mixture was loaded into a Noztek ProHT extruder to form the filament;  
the diameter of the obtained filament was controlled by an electronic caliper; 

\item finally, scintillating filament with a diameter of 1.75 $\pm$ 0.05~mm was produced.

\end{itemize}

The optimal temperature regimes for obtaining the scintillation filaments with the
addition of inorganic granules are shown in Table~\ref{tab:table1}.

\begin{table}[h!]
\begin{center}
\begin{tabular}{ |c|c| } 
\hline
 Type of polymer & Optimal printing temperature, $^o$C  \\
  \hline
  PS  & 200 - 235 \\ 
 \hline
  PMMA  & 250 - 270 \\ 
 \hline
  ABS  & 220 - 250 \\ 
 \hline
  SBS  & 210 - 230 \\ 
 \hline
\end{tabular}
\caption{
  \label{tab:table1} Temperature regimes for obtaining scintillation filaments depending on the type of
polymer pellets (binder)}
\end{center}
\end{table}

\begin{figure}[h]
\centering
{\includegraphics[width=9cm]{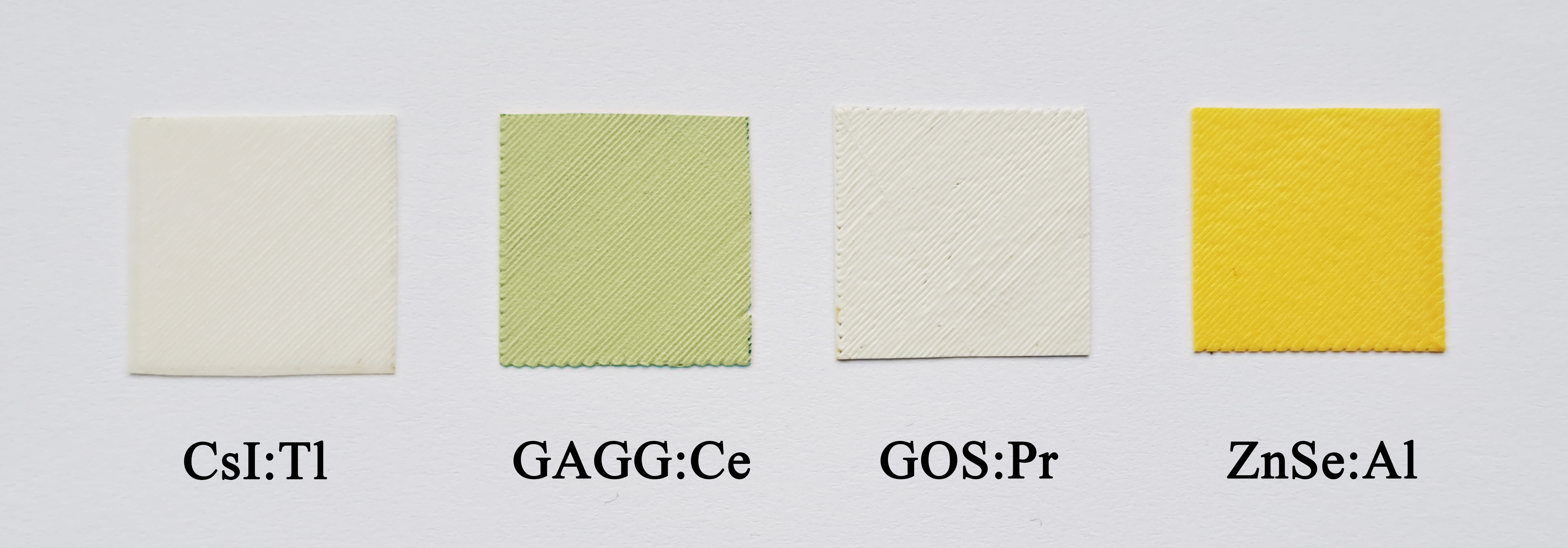}}
\caption{\label{fig:figure3} 3D printed samples $20\times20\times1$~mm$^3$ with inorganic scintillation granules CsI:Tl, GAGG:Ce, GOS:Pr, ZnSe:Al.}
\end{figure}

Scintillation filaments based on ZnSe:Al, GOS:Pr,  GAGG:Ce and CsI:Tl granules for 3D printing using
FDM technology have been developed. The FDM Creatbot F430 printer was used to obtain 3D printed samples.
SolidWorks 3D CAD software was used to design 3D scintillator sample model.
CreateWare was used as a slicer to break up the designed model into layers for 3D printing.
The printing bed and nozzle temperature as well as the printing speed and other parameters
were set depending on the type of filament. As we discovered earlier~\cite{Berns}, increasing the
temperature of chamber, bed and extruder improves the scintillator transparency.
The samples in this work were printed at the maximum printing temperature for each type of
polymer pellets, see Table~\ref{tab:table1}. The bed temperature at printing was 110 degrees, which is the
maximum temperature that the printer can handle. The printing speed was between 25~mm/s to 35~mm/s,
with a layer thickness between 0.1 to 0.2~mm. Figure~\ref{fig:figure3} shows 3D printed samples with the different
inorganic scintillation granules.

\section{Properties of the 3D printed inorganic scintillators}

\subsection{Research methodology}

Measurements of decay times were carried out by using a combined fluorescent lifetime and steady-state spectrometer
FLS 920 (Edinburgh Instruments) equipped with a hydrogen filled nF 900~ns flashlamp with optical pulse duration of
1.0 – 1.6~ns and pulse repetition rate of 40~kHz, for time correlated single photon counting measurements. 
The X-ray tube and the photodetector Hamamatsu R1926A were placed on the same side from the samples.
X-ray luminescence spectra were measured in the reflection mode under a 
steady-state X-ray excitation by applying voltage of 40~kV and current of 40~$\mu$A on Ag anode.
The emitted light was dispersed with a monochromator with grating 1200~grooves/mm.
The obtained spectra were not corrected for the spectral sensitivity of the detection system.

Measurements of the X-ray relative light output of the 3D printed scintillators were carried out on the stand
by measuring and comparing the light output of different scintillators~\cite{Tretyak}. The samples were irradiated with X-rays
at a source anode voltage of 90~kV. The radiation beam was limited by a lead collimator 3~mm in diameter.
The light output of the objects under the test was compared with the reference – a ZnSe:Al single crystal  $20\times20\times1$~mm$^3$ in size.
The scanning step was determined to be $5\times5$ mm$^2$. For one measuring process a reference sample and five 3D-printed samples
were located into the stand. With the selected scanning parameters, 16 measurement points on the samples can be obtained.
The measurement results were recorded in arbitrary units (values of the ADC stand). After averaging the values for each sample,
the light output of each sample relative to the reference was calculated using a proportional transformation. For example,
the signal of the reference sample is 987 and the signal of the 3D-printed sample is 546, then the relative
light output of the 3D-printed sample equals 100*546/987 = 55.32\%.
The scintillation light output efficiency spectra were measured with the Hamamatsu R1307 PMT by exciting
with alpha particles from Pu-239 source and beta particles from Bi-207 source. A signal from PMT anode was fed into Rigol DS1302CA oscilloscope.

The spatial resolution of 3D printed scintillators was determined using a self-made X-ray imaging setup~\cite{Galkin}
with the following characteristics: Isovolt Titan E 160 X-ray source with W anode; inherent filtration (mm) 0.8/Be;
max. Ua = 160~kV, Ia = 10~mA; nom. focal spot value IEC 336–1.5 (0.4); focal spot size EN 12 543 (mm) – 3(1).
Spatial resolution was determined on images taken with a digital photocamera by checking the quantity of
resolved wire pairs using an EN 462-5 Duplex IQ standard containing 13 pairs of wires with the different diameters.
The images were captured by a Canon camera with the $3264\times2448$ (8 megapixels) resolution. Distance from the X-ray
source to the composite scintillator was around 0.5~m. 
All the measurements were performed at room temperature.

\subsection{Studying luminescence and decay time of the samples}

\begin{figure}[h]
\centering
{\includegraphics[width=11cm]{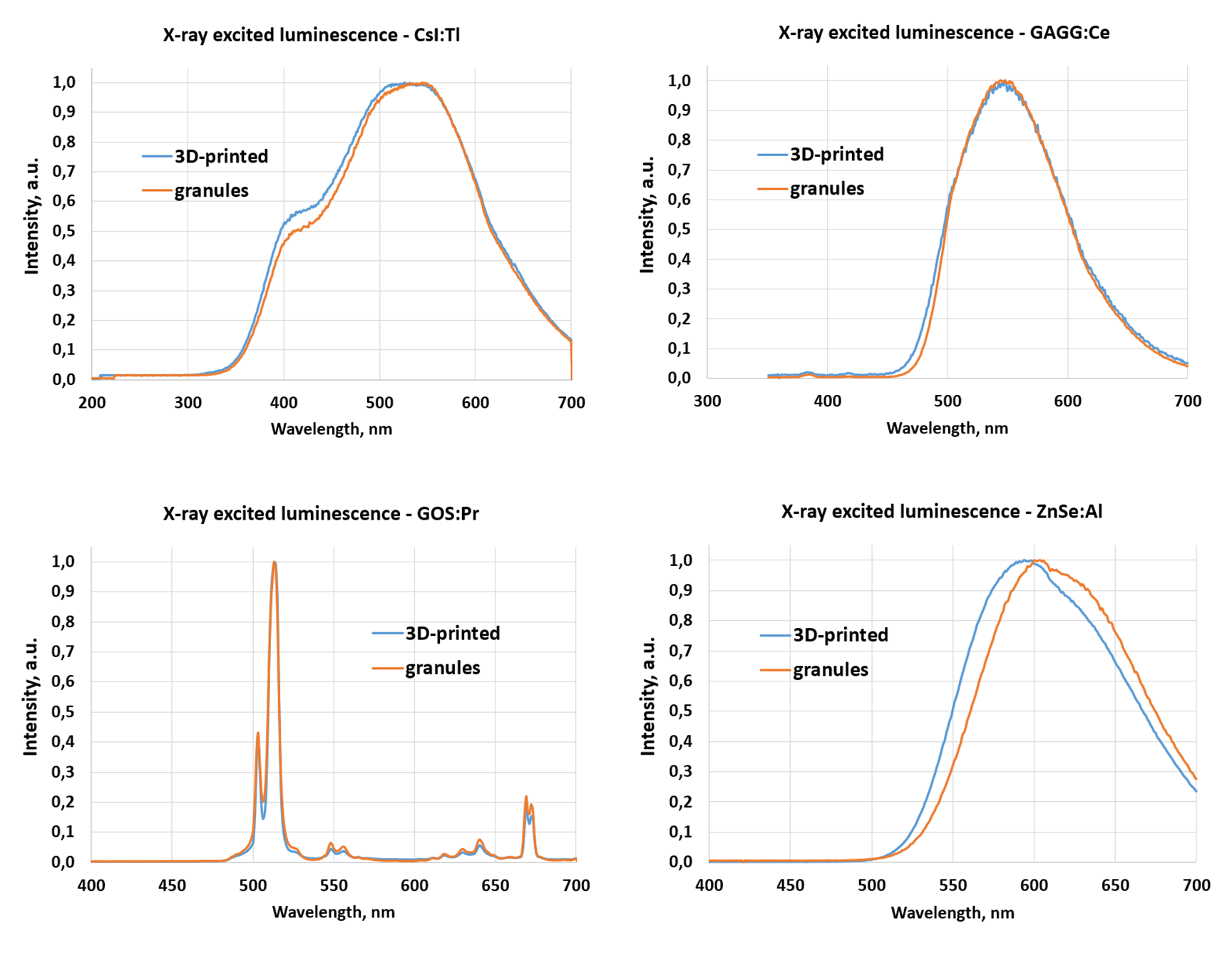}}
\caption{\label{fig:figure4} X-ray luminescence spectra of 3D printed samples and inorganic scintillation granules CsI:Tl, GAGG:Ce, GOS:Pr, ZnSe:Al.}
\end{figure}

\begin{figure}[h]
\centering
{\includegraphics[width=15cm]{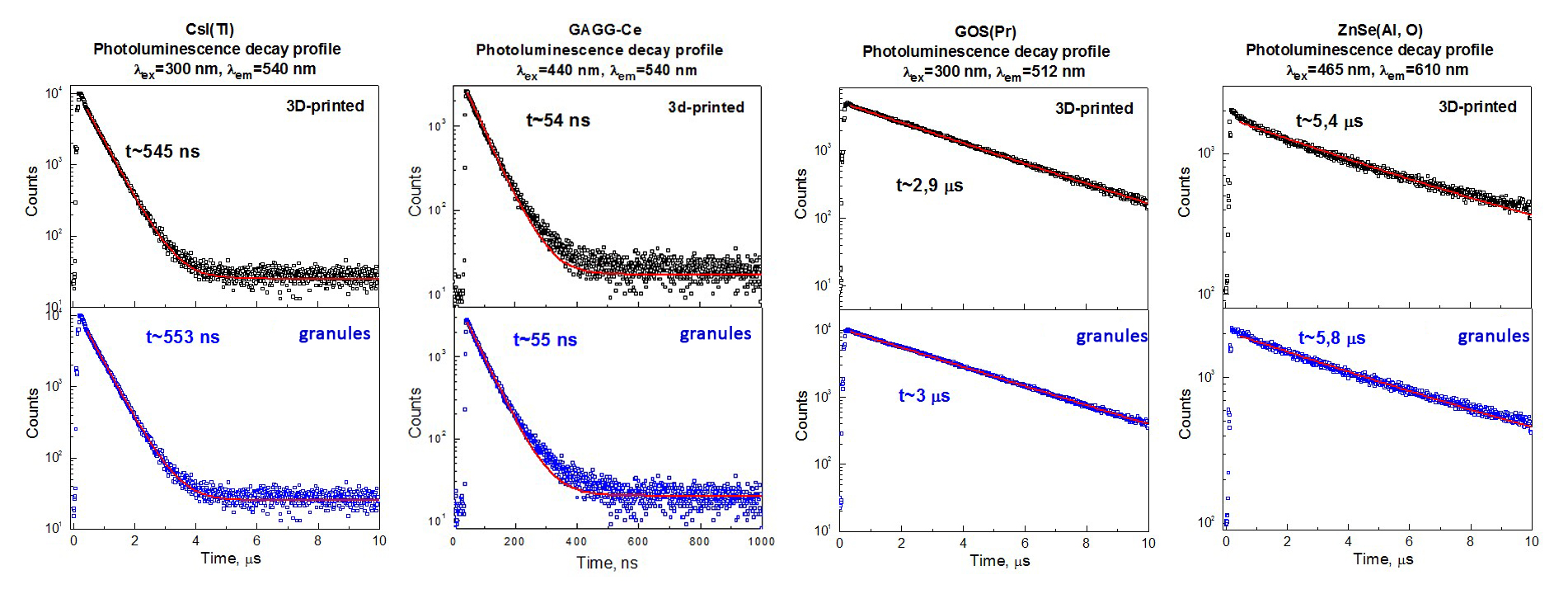}}
\caption{\label{fig:figure5} Photoluminescence decay profiles of 3D printed samples and inorganic scintillation granules CsI:Tl, GAGG:Ce, GOS:Pr, ZnSe:Al.}
\end{figure}

The X-ray luminescence and decay time results were compared with the inorganic scintillation granules obtained in Section~\ref{sec:second}
(see Figure~\ref{fig:figure1}).
The results of the study of luminescence and decay time of the 3D printed samples are shown at Figures~\ref{fig:figure4} and~\ref{fig:figure5}.
The scintillation properties of the 3D printed samples are similar to inorganic scintillator granules.
The size of the 3D printed samples was $20\times20\times1$mm$^3$. Inorganic scintillator granules for measurement were placed in $\oslash10\times2$ mm$^2$
cuvettes.

Thus, 3D printing, as well as the binder (polystyrene), does not lead to changes in the X-ray luminescence spectra and does not affect the decay time.

\subsection{X-ray registration: dependence of light output on particle size distribution, granules concentration and thickness and X-ray light output uniformity}
\begin{figure}[h]
\centering
{\includegraphics[width=7cm]{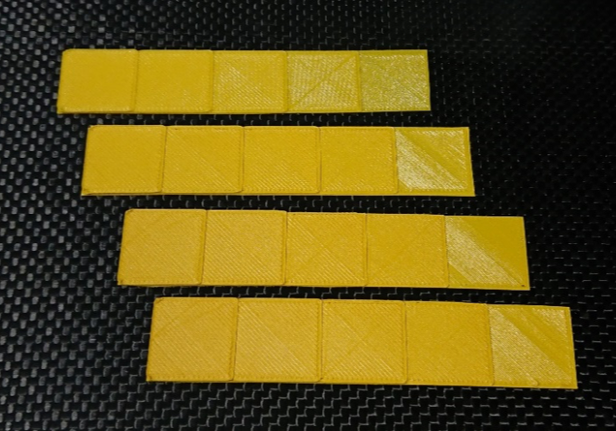}}
\caption{\label{fig:figure6} 3D printed scintillators with ZnSe:Al granules.}
\end{figure}

The samples of 3D printed scintillators with 25-75\% weight content of ZnSe:Al for X-ray registration were prepared as shown Figure~\ref{fig:figure6}.
Several thicknesses from 0.2~mm to 1.0~mm and granules size in the ranges 1 - 15, 45 - 63, 63 - 100 and 100 - 140~$\mu$m were tested.

\begin{figure}[h]
\centering
{\includegraphics[width=5.5cm]{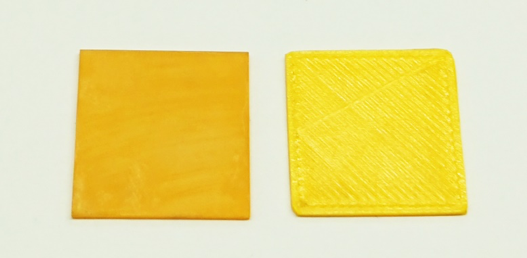}}
\caption{\label{fig:figure7} Left: ZnSe:Al single crystal. Right: 3D printed scintillator with ZnSe:Al granules.}
\end{figure}

As a reference for measurement, we used a ZnSe:Al single crystal with a size of $20 \times 20 \times 1$~mm$^3$ as shown Figure~\ref{fig:figure7}.

\begin{figure}[h]
\centering
{\includegraphics[height=6.7cm]{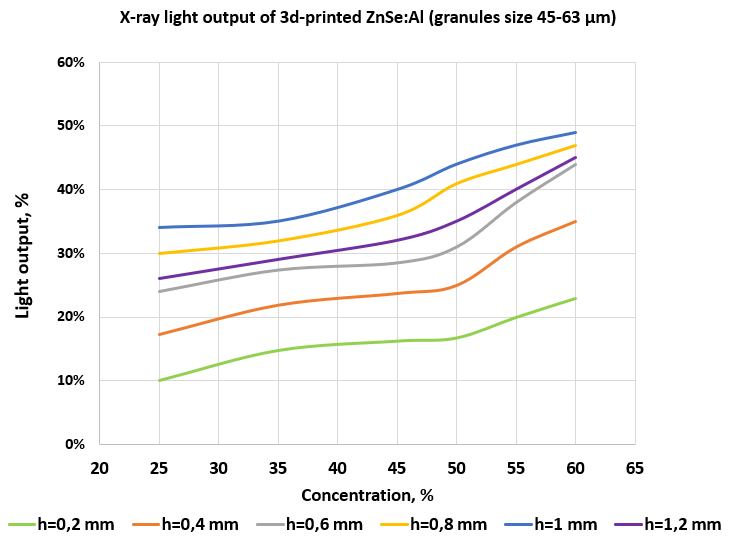}}
\caption{\label{fig:figure8} X-ray light output of 3D printed scintillators with different concentration of ZnSe:Al
  granules relative to the single crystal ZnSe:Al. Different sample thicknesses (h) were tested (granules size was 45-63 micrometers).}
\end{figure}

The dependence of the light output on the concentration of the inorganic scintillator granules is shown in Figure~\ref{fig:figure8}.
The concentration of the inorganic granules in the filament is shown on the X axis while the light output is given on the Y axis.
Samples of various thicknesses, from 0.2 to 1.2~mm, were tested and compared to the industrial scintillator sample.

\begin{figure}[h]
\centering
{\includegraphics[height=6.7cm]{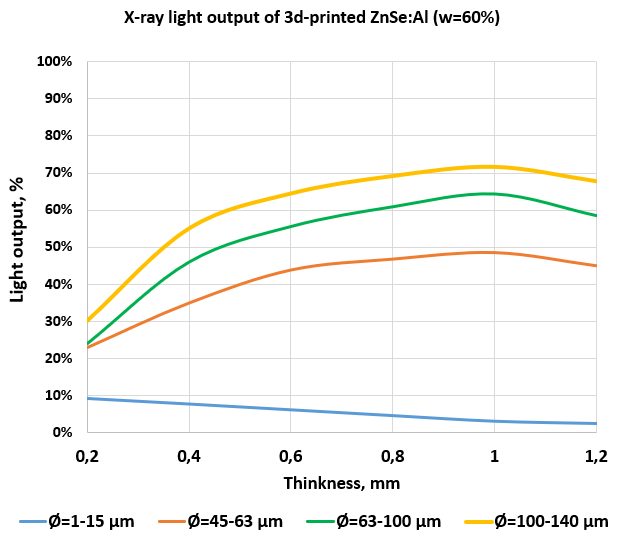}}
\caption{\label{fig:figure9} X-ray light output of 3D printed scintillators with different thickness relative to the single crystal ZnSe:Al $20 \times 20 \times 1$~mm$^3$. Different granules size (Ø) were tested (weight content of inorganic granules was 60\%).}
\end{figure}

The optimal weight content of inorganic granules in the scintillation filament is obtained to be 50-60\%. Printing filament with the granules
weight content more than 60\% is difficult, as the filament crumbles. Printing filament with a granule weight content less than 50\%
leads the light output decrease, thus a lower particle registration efficiency.
The dependence of the light output on the sample thickness is shown in Figure~\ref{fig:figure9}. The optimal sample thickness is obtained to be 0.8-1.0~mm.

\begin{figure}[h]
\centering
{\includegraphics[width=9.5cm]{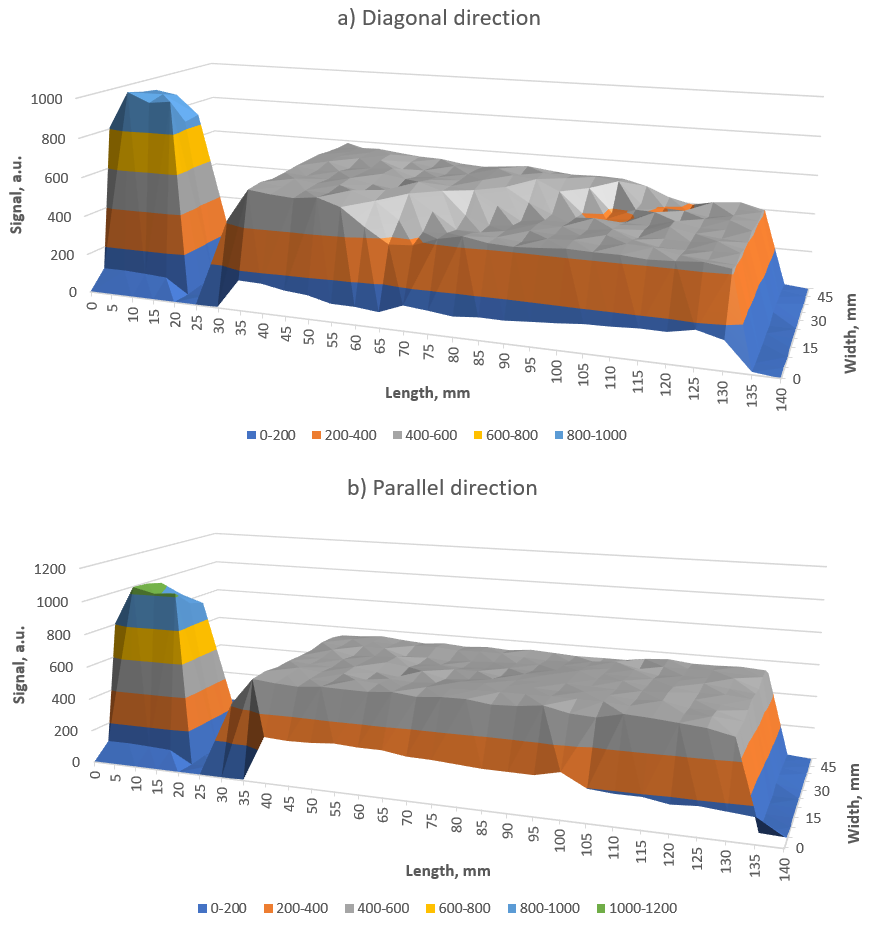}}
\caption{\label{fig:figure10} X-ray light output uniformity: a) sample with the diagonal direction of the nozzle stroke; b) sample with the parallel direction of the nozzle stroke.}
\end{figure}

To study the uniformity of the distribution of inorganic granules inside the filament, a sample of $60 \times 100 \times 0.3$~mm$^2$ was printed.
A filament with ZnSe:Al granules 63 - 100~$\mu$m in size and 30\% weight content was used for printing. The X-ray light output
uniformity of this was studied. As a reference was used a ZnSe:Al single crystal with a size of $20 \times 20 \times 0.3$~mm$^3$, which was
located near the 3D-printed sample for measuring. The visualization of the signals obtained on this sample is shown in Figure~\ref{fig:figure10}-a.
The nonuniformity of the light output of the sample was defined to be the ratio of signals: (maximum-minimum)/ average, is determined to be 46\%. However, the visualization of the signal shows that such a high nonuniformity is
associated with the direction of the nozzle stroke. Therefore, it was decided to change the direction of the nozzle stroke.
To do this, in the CreatWare program, the position 
of the sample on the stage was shifted by 45 degrees. Thus, instead of the diagonal direction of the nozzle stroke,
we got a parallel one. In this way, a second sample of the same size was printed using the same filament.
The visualization of the signals obtained on the second sample is shown in Figure~\ref{fig:figure10}-b.
The nonuniformity of the second sample was 15\%, which is probably due to the nonuniformity of the distribution of inorganic granules inside the filament.

\subsection{Registration of alpha- and beta-radiation: dependence of light output on polymers as optical binder}

\begin{figure}[h]
\centering
{\includegraphics[width=13cm]{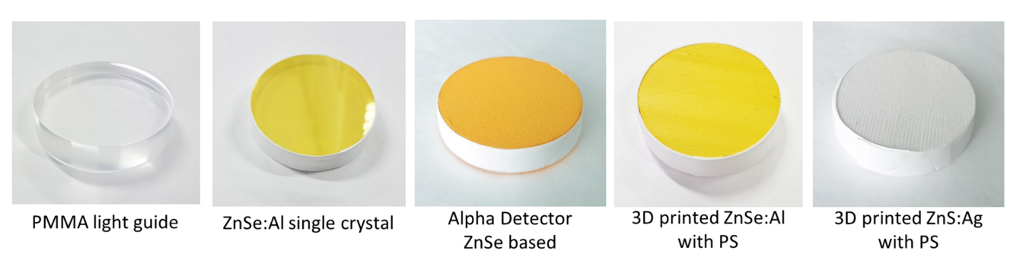}}
\caption{\label{fig:figure11} PMMA light guide and samples for alpha particle registration.}
\end{figure}

Two inorganic scintillators with ZnSe:Al and ZnS:Ag were 3D printed for detection of alpha particles.
The size of the inorganic granules was 40 - 63~$\mu$m, the size
of 3D printed samples was $\oslash 44 \times 0.2$~mm$^2$. The samples based on ZnSe:Al granules were printed with PS,
ABS, SBS, and PMMA polymers as optical binder.
The samples based on ZnS:Ag granules were printed only with PS polymers as optical binder. As a reference, the $\oslash44 \times 1$~mm$^2$
ZnSe:Al single crystal produced by ISMA~\cite{isma} was used. Both the 3D printed samples and ZnSe:Al single crystal were coupled to
$\oslash 44 \times 8$~mm$^2$ PMMA light guide (see Figure~\ref{fig:figure11}) for a better light collection. The side surface of the
light guides is wrapped in two layers
of a PTFE membrane. Also, as references Ø44 mm Alpha Detector ZnSe based by ISMA~\cite{isma} was used, this sample already contains $\oslash44 \times 8$~mm$^2$
PMMA light guide in its design. The samples are shown in Figure~\ref{fig:figure11}.

Each sample was exposed with a Pu-239 source up to a set of 100 thousand events. The light output was determined from the maximum of
the amplitude spectrum. Among various polymers the best results were obtained with PS and PMMA as optical binder (see Table~\ref{tab:table2}).
Their light output reaches 94\% compared to ZnSe:Al single crystal. However, it should be noted that the counting rate
of these 3D printed samples was 2.5 times lower compared to the single crystal. The counting rate was determined
as the ratio of the number of events to the exposure time. This means that it took 2.5 times longer to set
100 thousand events for a 3D-printed sample. The decrease in the counting rate may be related to a decrease in the
probability of interaction of alpha particles with a scintillation material, since some of the alpha particles will be absorbed by the optical
binder without light emission. It also can be due to the absorption and scattering of light in the volume of the 3D printed scintillator.

\begin{figure}[h]
\centering
{\includegraphics[width=8cm]{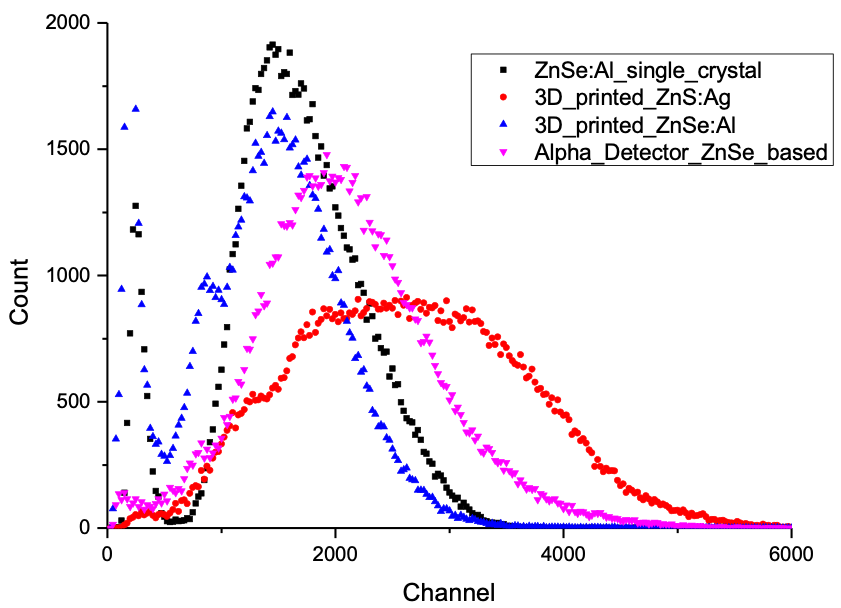}}
\caption{\label{fig:figure12} Spectra of registration of alpha-radiation (Pu-239) of ZnSe:Al and ZnS:Ag 3D
  printed samples with PS as optical binder, ZnSe:Al single crystal and Alpha Detector ZnSe based.}
\end{figure}

The ZnS:Ag 3D printed sample shows a very good result, its relative light output is 180\% and counting rate is 100\% compared to ZnSe:Al single crystal.
Spectra of registration of alpha-radiation from Pu-239 are shown on the Figure~\ref{fig:figure12}.

\begin{table}[h!]
\begin{center}
\begin{tabular}{ |c|c|c| } 
  \hline
 Sample & Relative light output, \% & Counting rate, \%   \\
       
 \hline
ZnSe:Al single crystal  	& 100	& 100 \\\hline
Alpha Detector ZnSe based	& 127	& 200 \\\hline
3D printed ZnS:Ag with PS	& 180	& 100 \\\hline
3D printed ZnSe:Al with PS	&  94	&  40 \\\hline
3D printed ZnSe:Al with PMMA	&  94	&  40 \\\hline
3D printed ZnSe:Al with ABS	&  66	&  20 \\\hline
3D printed ZnSe:Al with SBS	&  39	&  10 \\\hline
\end{tabular}
\caption{
  \label{tab:table2} Relative light output of 3D printed ZnSe:Al and ZnS:Ag based scintillators with different plastics
  as optical binder under Pu-239 source (weight content of inorganic granules was 50\%)}
\end{center}
\end{table}

\begin{figure}[h]
\centering
{\includegraphics[width=9cm]{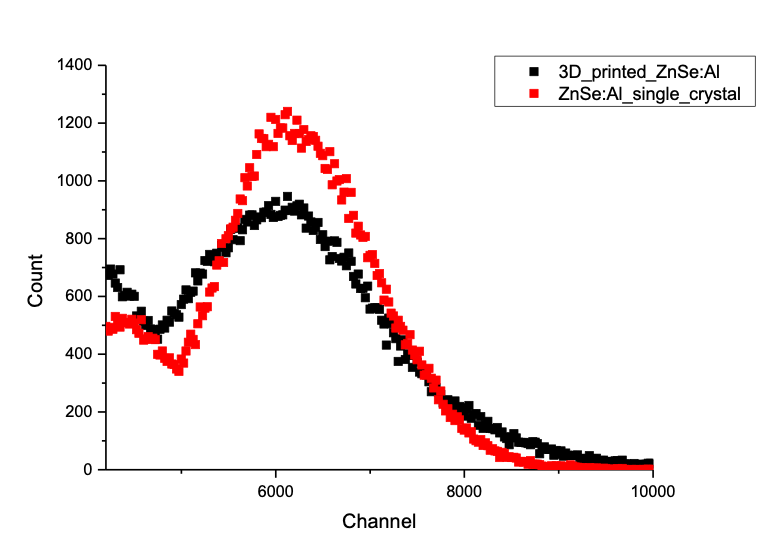}}
\caption{\label{fig:figure13} Spectra of registration of beta-radiation (Bi-207) of ZnSe:Al 3D printed
  sample with PS as optical binder and ZnSe:Al single crystal}
\end{figure}

3D printed sample with size $20 \times 20 \times 1$~mm$^3$ was printed for detection of beta-particles. The size of ZnSe:Al granules
was 100 - 300~$\mu$m.
3D printed samples with a granules size of <100~$\mu$m turned out to be insensitive to beta-radiation.
Producing a filament and 3D printing with granules >300~$\mu$m is difficult due to the rapid wear of extruder parts,
as well as nozzles and feed rollers of the 3D printer. Therefore, the granule size of 100 - 300~$\mu$m is optimal for 3D
printing of samples sensitive to beta-radiation. As a reference, $20 \times 20 \times 1$~mm$^3$ ZnSe:Al single crystal was used. 
For beta-irradiation the light output was determined from the position of the maximum of the total absorption peak of
conversion electrons with energy of 976~keV. The light output of the 3D printed sample was 98\% compared to the single crystal.
But as in the case of irradiation with alpha-particles, the counting rate of the printed sample
was also 2.5 times lower than that of the single crystal. The spectra of the registration of beta-radiation (Bi-207) are shown in Figure~\ref{fig:figure13}.

\begin{table}[h!]
\begin{center}
\begin{tabular}{ |c|c|c| } 
  \hline
 Sample & Relative light output, \% & Counting rate, \%   \\
        
 \hline
ZnSe:Al single crystal  	& 100	& 100 \\\hline
3D printed ZnSe:Al with PS	&  98	&  40 \\\hline
\end{tabular}
\caption{
  \label{tab:table3} Relative light output of 3D printed ZnSe:Al based scintillator with PS as optical binder under
  Bi-207 source (weigh content of inorganic granules was 50\%)}
\end{center}
\end{table}

\begin{figure}[h]
\centering
{\includegraphics[width=8cm]{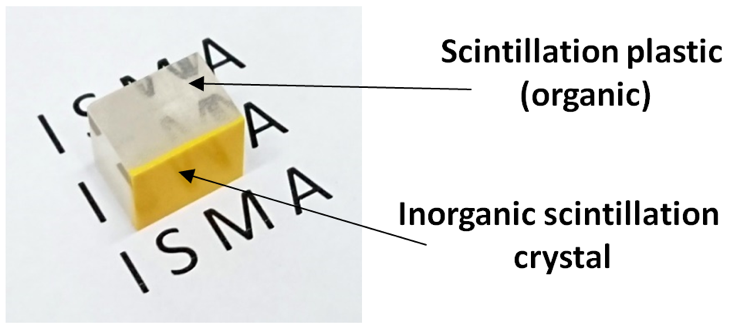}}
\caption{\label{fig:Figure14} First 3D printed prototype of the combined detector after post-processing.}
\end{figure}

3D printing allows to print both organic and inorganic scintillators for simultaneous alpha- and beta- registration. There can be prospects for combining organic and inorganic scintillators in a detector with increased registration sensitivity. We have printed the first prototype of the combined detector, which is shown in Figure~\ref{fig:Figure14}. Combinations of different scintillators result in achieving unique effects and this will be the subject of further research.

\subsection{X-ray imaging}
\begin{figure}[h]
\centering
{\includegraphics[width=15cm]{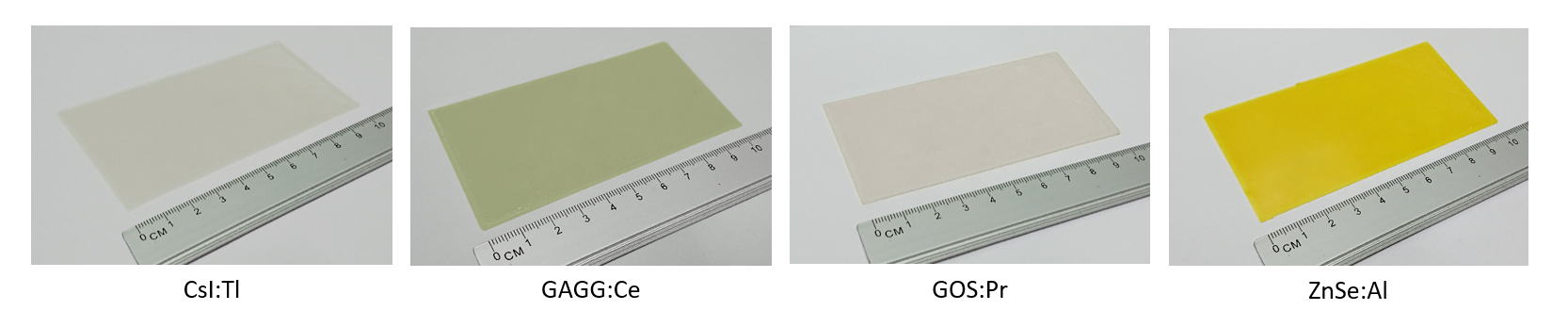}}
\caption{\label{fig:Figure15} 3D printed samples for X-ray imaging.}
\end{figure}

Films for X-ray imaging, effective for registration of soft X-ray radiation (20 - 90~keV),
were obtained on the basis of ZnSe:Al, GOS:Pr,  GAGG:Ce and CsI:Tl granules (see Figure~\ref{fig:Figure15}).
The size of granules was 1 - 15~$\mu$m. Composite filaments were fabricated with 60\% by weight of
scintillation granules using PMMA as a binder 
medium. Composite films were obtained with a thickness of 0.15~mm to 0.3~mm.
The area of the films is $100 \times 45$~mm$^2$. 

\begin{figure}[h]
\centering
{\includegraphics[width=15cm]{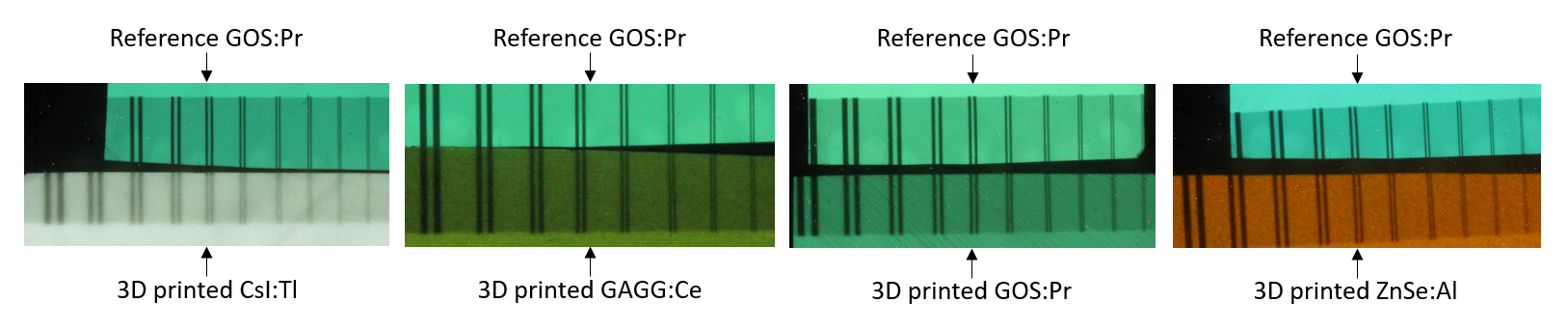}}
\caption{\label{fig:Figure16} X-ray imaging screens of 3D printed films.}
\end{figure}

As a reference, composite samples based on GOS:Pr
(size of granules was 1 - 5~$\mu$m), size of $60 \times 40 \times 0.10$~mm$^3$ was used. Organic polysiloxane matrix was prepared as a reference, as shown in~\cite{Gerasymov}. The best results for 3D printed samples were obtained at a film thickness of 0.15~mm. Spatial resolution for these 3D printed films were from 3.15 line pairs per mm to 3.35 line pairs per mm (Table~\ref{tab:table4}). 
X-ray imaging screens of 3D printed films are shown in Figure~\ref{fig:Figure16}. The radiation resistance of 3D printed samples will be studied in further research.

\begin{table}[h!]
\begin{center}
\begin{tabular}{ |c|c|c|c|c|c| } 
  \hline
Scintillator & Reference GOS:Pr & ZnSe:Al & GOS:Pr & GAGG:Ce & CsI:Tl \\\hline
Spatial resolution & \multirow{2}{*}{3.85} & \multirow{2}{*}{3.15} & \multirow{2}{*}{3.35} &  \multirow{2}{*}{3.35} &  \multirow{2}{*}{3.15} \\
line pairs per mm  &      &      &      &      &       \\\hline 
\end{tabular}
\caption{
  \label{tab:table4} Spatial resolution of 3D printed composite films 0.15~mm thick}
\end{center}
\end{table}

\newpage

\section{Conclusions}

Using 3D printing, it is possible to manufacture large area detectors. There is no need
for post-processing: the surface is ready for connection with the PMT or SiPM. It is cost
effective due to the possibility of using the waste material from the production of single crystals. 

The developed scintillators can be used for registration of X-ray radiation and for
the detection of alpha and beta particles. Moreover, the developed scintillators can be
used in high-energy physics. For example, 3D printing technology will permit the easier development of finely segmented EM calorimeters.

Further development of 3D printing of scintillators can be directed to the creation of
radiation hardness scintillators and the creation of combined detectors.


\begin{thebibliography}{99}

\bibitem{3DETwebsite} The 3D Printed Detector (3DET) project, \url{https://threedet.web.cern.ch}.

\bibitem{Berns} S. Berns, A. Boyarintsev, S. Hugon, U. Kose, D. Sgalaberna, A. De Roeck, A. Lebedynskiy, T. Sibilieva, and P. Zhmurin. A novel polystyrene-based scintillator production process involving additive manufacturing. JINST \textbf{15} (2020) no.10, 10, 
doi:10.1088/1748-0221/15/10/P10019

\bibitem{3Dmatrix} The 3DET collaboration, S. Berns, E. Boillat, A. Boyarintsev,  A. De Roeck, S. Dolan, A. Gendotti, B. Grynyov, S. Hugon, U. Kose, S. Kovalchuk, A. Rubbia, T. Sibilieva, D. Sgalaberna, T. Weber, J. Wuthrich, X. Y. Zhao. Additive manufacturing of fine-granularity optically-isolated plastic scintillator elements. JINST \textbf{17} (2022) no.10, P10045, doi:10.1088/1748-0221/17/10/P10045

\bibitem{Blondel}
A. Blondel et al., A fully active fine grained detector with three readout views, 2018 JINST 13 P02006
[arXiv:1707.01785].

\bibitem{Dujardin}
C.~Dujardin, E.~Auffray, E.~Bourret-Courchesne, P.~Dorenbos, P.~Lecoq, M.~Nikl, A.~N.~Vasil'ev, A.~Yoshikawa and R.~Y.~Zhu,
``Needs, Trends, and Advances in Inorganic Scintillators,''
IEEE Trans. Nucl. Sci. \textbf{65} (2018) no.8, 1977-1997, 
doi:10.1109/TNS.2018.2840160


\bibitem{Auffray} E. Auffray. Scintillation properties and radiation hardness of the Ce- doped, Codoped Ce, Mg garnet crystals and garnet crystal fibers development, LHCB upgrade calorimeter, available at \url{https://indico.cern.ch/event/706303/contributions/2898330/attachments/1605966/2548030/GarnetpropertiesEauffray23022018.pdf }(accessed on August 18th, 2022)

\bibitem{Martinazzoli} L. Martinazzoli, N. Kratochwil, S. Gundacker, E. Auffray. Scintillation properties and timing performance of state-of-the-art Gd3Al2Ga3O12 single crystals, Nuclear Inst. and Methods in Physics Research, A, 1000 (2021) 165231, \url{https://doi.org/10.1016/j.nima.2021.165231}

\bibitem{Martinazzoli_2} L. Martinazzoli, N. Kratochwil, S. Gundacker, E. Auffray. Formation of color centers at X-ray irradiation of ZnSe single crystals, Radiation Measurements, (2020) 106232, \url{https://doi.org/10.1016/j.nima.2021.165231}



  
  



\bibitem{Gerasymov} 
I.~Gerasymov, T.~Nepokupnaya, A.~Boyarintsev, 
O.~Sidletskiy, D.~Kurtsev, O.~Voloshyna, O.~Trubaieva, Y.~Boyarintseva,
T.~Sibilieva, A.~Shaposhnyk, 
O.~Opolonin, S.~Tretyak. 
GAGG:Ce composite scintillator for X-ray imaging, 
Optical Materials 109, 2020, 110305, doi:10.1016/j.optmat.2020.110305

\bibitem{Boyarintsev} 
A. Boyarintsev, N. Galunov, I. Gerasymov, N. Karavaeva, A. Krech, L. Levchuk, V. Popov, O. Sidletskiy, P. Sorokin, O. A. Tarasenko. Radiation-resistant composite scintillators based on GSO and GPS grains, Nuclear Instruments and Methods in Physics Research Section A-accelerators Spectrometers Detectors and Associated Equipment, 2017. doi:10.1016/J.NIMA.2016.10.034

\bibitem{Boyarntsev_2} A. Boyarintsev, N. Galunov, I. Gerasymov, T. Gorbacheva, B. Grinyov, N. Karavaeva, S. U. Khabuseva, A. Krech, L. Levchuk, E. Martynenko, V. Popov, O. Sidletskiy, O. A. Tarasenko. Radiation Resistance of Composite Scintillators, Problems of Atomic Science and Technology, 2019. doi:10.46813/2019-121-060

\bibitem{Kerch} A. Krech, N. Galunov, A. Boyarintsev, et al., Radiation Resistance of Composite Scintillators Based on Grains of Oxide Single Crystals, Acta Physica Polonica A., 141 (4) (2022) 426-435.

\bibitem{Nepokupnaya} T. Nepokupnaya, A. Boyarintsev, B. Grynyov, S. Galkin, A. Kolesnikov, The ways of characteristics improvement for alpha–beta detectors on ZnSe based composite scintillators, Nuclear Instruments \& Methods in Physics Research Section A-accelerators Spectrometers Detectors and Associated Equipment, (2021) 165704. 

\bibitem{Tretyak} A measuring complex for control the uniformity of the light output of scintillators / S.O. Tretyak, O.V. Popkova // Functional Materials 25 (4), 2018, p. 835-837

\bibitem{Galkin} S.M. Galkin, I.A. Rybalka, I.A. Tupitsyna, V.S. Zvereva, V.A. Litichevskiy, The development of flexible scintillation panels based on chalcogenide and oxide phosphors for advanced x-ray scanners and tomographs, Science and Innovation. 12 (6) (2016) 37–45, \url{doi: https://doi.org/10.15407/scine12.06.037}

\bibitem{Barsuk} Barsuk for the LHCb Calorimeter Group, “The Shashlik electromagnetic calorimeter for the LHCb experiment”, LHCb-PROC-2004-006, 11th Intern. Conf. On Calorimetry In HEP, Perugia (Italy) 2004

\bibitem{isma} \url{http://www.isma.kharkov.ua/en/node/31}


\end{thebibliography}
\end{document}